\documentclass[twocolumn,showpacs,superscriptaddress,prbx]{revtex4}

\usepackage{amsmath,amssymb,graphicx}
\usepackage{color}
\usepackage{xspace}

\newcommand{\oh}{\mbox{$\frac{1}{2}$}}

\newcommand{\las}[0]{\langle}
\newcommand{\ras}[0]{\rangle}

\newcommand{\eg}[0]{e.g.\@\xspace}

\newcommand{\UP}[0]{\uparrow}

\newcommand{\si}[0]{\sigma}
\newcommand{\om}[0]{\omega}

\newcommand{\kF}{k_\text{F}}
\newcommand{\vF}{v_\text{F}}

\newcommand{\nag}{{\phantom{\dag}}}

\begin{document}

\title{Luttinger Liquid Physics and Spin-Flip Scattering on Helical Edges}

\author{M. Hohenadler}

\affiliation{\mbox{Institut f\"ur Theoretische Physik und Astrophysik,
    Universit\"at W\"urzburg, Am Hubland, 97074 W\"urzburg, Germany}}

\author{F. F. Assaad}

\affiliation{\mbox{Institut f\"ur Theoretische Physik und Astrophysik,
    Universit\"at W\"urzburg, Am Hubland, 97074 W\"urzburg, Germany}}

\affiliation{Kavli Institute for Theoretical Physics, University of California, Santa Barbara, CA, 93106}

\begin{abstract}
  We investigate electronic correlation effects on edge states of quantum
  spin-Hall insulators within the Kane-Mele-Hubbard model by means of quantum
  Monte Carlo simulations. Given the $U(1)$ spin symmetry and time-reversal
  invariance, the low-energy theory is the helical Tomanaga-Luttinger model,
  with forward scattering only. For weak to intermediate
    interactions, this model correctly describes equal-time spin and charge
  correlations, including their doping dependence.  
  As apparent from the Drude weight, bulk states become relevant in the
  presence of electron-electron interactions, rendering the forward-scattering model 
  incomplete. Strong correlations give rise to
  slowly decaying transverse spin fluctuations, and inelastic spin-flip
  scattering strongly modifies the single-particle spectrum, leading to
  graphene-like edge state signatures. The helical Tomanaga-Luttinger model 
  is completely valid only asymptotically in the weak-coupling limit.
\end{abstract}

\date{\today}

% see
% http://www.aip.org/pacs/pacs2010/individuals/pacs2010_regular_edition/alpha_index.html
% 03.65.Vf Topological phases (quantum mechanics) 71.10.Pm Luttinger liquid
% 71.27.+a Strongly correlated electron systems; heavy fermions

\maketitle

{\it Introduction.}---A unique feature of quantum spin Hall insulators
(QSHIs), or two-dimensional (2D) topological insulators, are metallic edge
states with remarkable properties \cite{KaMe05a}. Contrary to chiral quantum
Hall edge states, QSHI edge states are helical, so that electrons with
opposite spin propagate in opposite directions. Due to time-reversal
invariance (TRI), the helical edge states are protected against disorder and
single-particle backscattering \cite{KaMe05a,Wu06}.  They are also
holographic in the sense that they exist only as edges of 2D systems
\cite{Wu06}, and can therefore not be completely separated from the bulk.
The number of pairs of edge states is directly related to the second
Chern number or $Z_2$ invariant \cite{KaMe05a}. For a review of topological 
insulators, see Ref.~\cite{HaKa10}.

Quantum fluctuations play a significant role for 2D topological insulators.
In particular, the one-dimensional (1D) edges have no well-defined
quasiparticle excitations, and are usually described using the framework of
bosonization or Luttinger liquid (LL) theory, which becomes particularly
simple in the presence of TRI and $U(1)$ spin symmetry. In this case
[referred to in the following as the helical Tomanaga-Luttinger (HTL) model],
only forward scattering is possible \cite{Giamarchi}.  {\it A priori}, such a
theory is only valid at low energies. Nevertheless, for metallic 1D systems,
LL theory provides a complete low-energy description even for strong
interactions. For helical edge states, the presence of bulk states is
intimately connected with the topological character of the system.  Strong
interactions on or beyond the size of the bulk band gap can give rise to a
substantial mixing of the different energy scales, and may explain deviations
of, \eg, the experimentally measured conductance \cite{Koenig07} from
expectations based on a low-energy description. Bulk effects fall outside the
regime of bosonization, and require a model which captures all relevant
energy scales. A comparison of the spectral properties of helical, spiral and
standard Luttinger liquids has been given in Ref.~\cite{PhysRevB.85.035136}.

The Kane-Mele (KM) model \cite{KaMe05a} of noninteracting electrons on the
honeycomb lattice with spin-orbit (SO) coupling $\lambda$ is a theoretical
framework to study $Z_2$ QSHIs. For small enough Rashba coupling, the ground
state for $\lambda>0$ is a topological band insulator (TBI).  The addition of
a Hubbard interaction term (KMH model) permits to study a strongly correlated
TBI \cite{RaHu10}, although it completely detaches the KM model from its
original motivation by graphene \cite{KaMe05b}. The electron-electron
interaction leads to a complex and rich many-body problem, which includes a
quantum spin liquid phase, and a magnetic transition at large Hubbard $U$
\cite{RaHu10,PhysRevLett.106.100403,PhysRevB.83.205122,Zh.Wu.Zh.11,Yu.Xie.Li.11,Wu.Ra.Li.LH.11,Ho.Me.La.We.Mu.As.12}.

In this Rapid Communication, using large-scale quantum Monte Carlo (QMC)
simulations of a previously introduced effective model
\cite{PhysRevLett.106.100403}, we provide a comprehensive assessment of the
validity of the HTL model. While confirming the interaction and doping
dependence predicted for a helical liquid in the weakly interacting limit, we
find significant deviations with increasing correlations which are beyond the
usual low-energy description and had remained unnoticed in previous numerical
work \cite{PhysRevLett.106.100403,Yu.Xie.Li.11,Zh.Wu.Zh.11}.  Additional
features in the single-particle spectrum are explained by inelastic spin-flip
scattering arising from magnetic fluctuations at the edge, driven by strong
electronic correlations. The interaction-driven mixing of multiple energy
scales is more subtle than the invalidation of the HTL model due to the
breaking of TRI, for example by means of strong bulk interactions
\cite{RaHu10,PhysRevLett.106.100403,Ho.Me.La.We.Mu.As.12,Zh.Wu.Zh.11} which
destroy the topological character, or by sufficient renormalization of the LL
parameter in the presence of Rashba coupling \cite{Wu06}.

{\it Model.}--- The phase diagram of the KMH model in the $\lambda$--$U$
plane is known from exact QMC simulations
\cite{PhysRevLett.106.100403,Zh.Wu.Zh.11,Ho.Me.La.We.Mu.As.12}; many of its
overall features are also captured by approximate methods
\cite{RaHu10,PhysRevB.83.205122,Yu.Xie.Li.11,Wu.Ra.Li.LH.11}. For
$U/t\lesssim3$ ($t$ being the hopping integral), the ground state is a TBI
with helical edges for any $\lambda>0$. The TBI with repulsive $U>0$ is
adiabatically connected to $U=0$
\cite{PhysRevLett.106.100403,Ho.Me.La.We.Mu.As.12}, suggesting that bulk
interactions are of minor importance in the TBI phase. Based on this result
and the fact that the edge states are exponentially localized at the edge
\cite{PhysRevLett.106.100403}, we have previously proposed an effective model
for the helical edge states (which exist throughout the TBI phase) with
Hubbard interaction only at one zigzag edge of a semi-infinite honeycomb
ribbon. It is defined by the action \cite{PhysRevLett.106.100403}
\begin{align}\nonumber \label{eq:action}
  \mathcal{S} = &-\sum_{\sigma,r,r'} \iint_{0}^{\beta} d\tau d\tau'
  c^{\dagger}_{r\sigma}(\tau) {G^\sigma_{0}}^{-1} (r-r',\tau-\tau')
  c^\nag_{r'\sigma}(\tau')
  \\
  &+ U \sum_{r} \int_{0}^{\beta} d\tau \left[n_{r\uparrow}(\tau) -\oh\right]
  \left[n_{r\downarrow}(\tau) - \oh\right]\,,
\end{align}
where $r$ numbers sites on the zigzag edge, and $G^\si_{0}$ is the Green's
function of the KM model through which the noninteracting bulk is taken into
account.

{\it Method.}---Equation~(\ref{eq:action}) corresponds to a 1D problem with a
bath that can be solved exactly using the continuous-time QMC (CTQMC) method
\cite{Rubtsov05,PhysRevLett.106.100403} on large systems at low temperatures.
Two crucial methodological developments compared to
Ref.~\onlinecite{PhysRevLett.106.100403} are the extension to a projective
zero-temperature scheme with projection parameter $\theta$, and to
grand-canonical simulations away from half filling ($n=1$). These advances
allow for a quantitative test of LL theory. We use the ribbon geometry of
\cite{PhysRevLett.106.100403}, with dimensions $L\times L'$ ($L'=64$), and
periodic (open) boundaries in the $x$ ($y$) direction. The hopping integral
and lattice constant are set to 1. The ratio $U/\lambda$ of the remaining two
parameters controls the degree of correlations. To study strong interactions
inside the TBI phase of the KMH model, we take $U=2$ and vary $\lambda$
\cite{PhysRevLett.106.100403,Ho.Me.La.We.Mu.As.12}.

\begin{figure}[t]
  \includegraphics[width=0.45\textwidth]{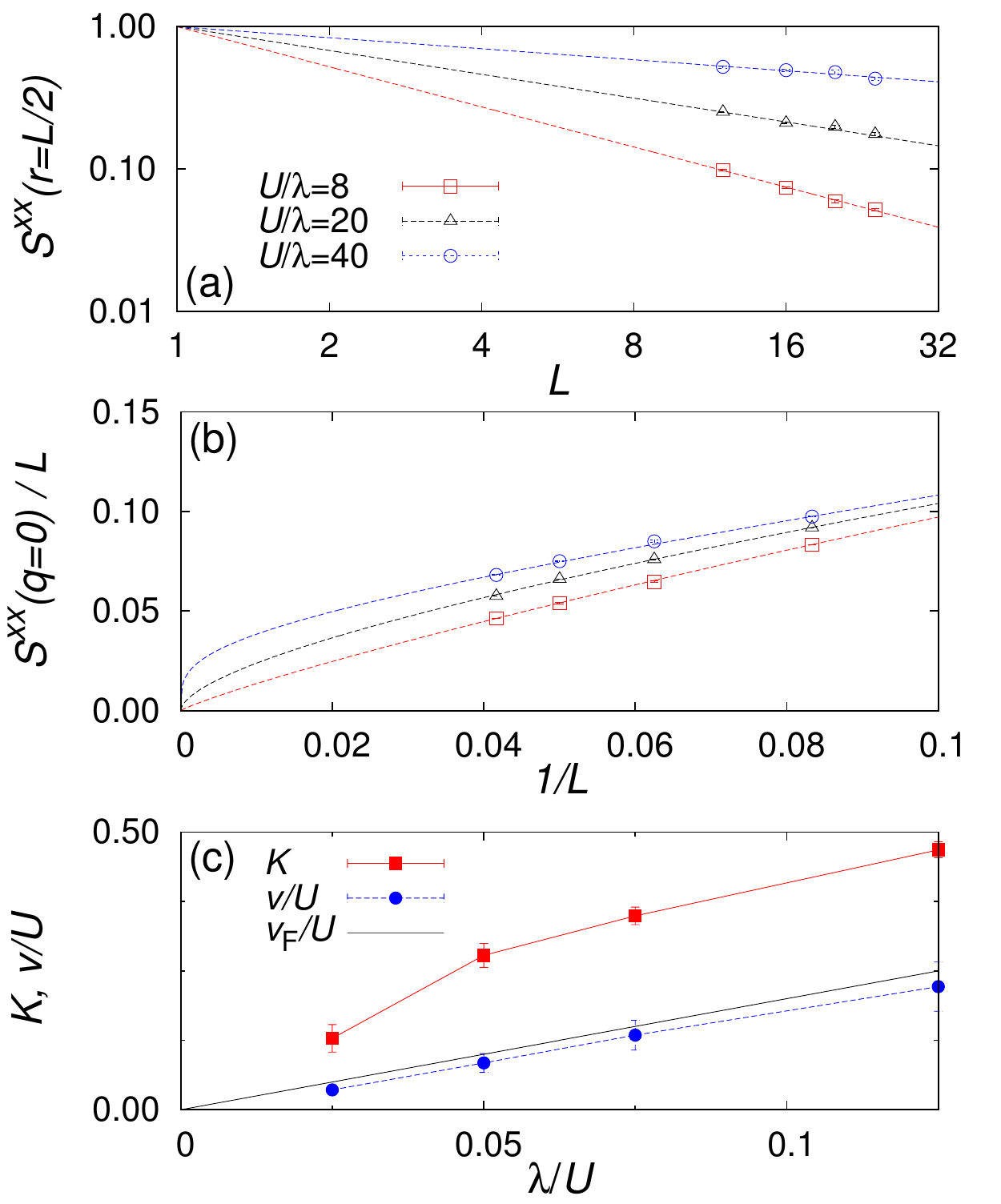}
  \caption{\label{Equal-time-correlations.fig} (Color online) Projective
    CTQMC results for the effective model~(\ref{eq:action}).  (a) Real-space
    correlations $S^{xx}(r=L/2)$ as a function of system size $L$ at
    different $U/\lambda$, normalized to 1 at $r=1$ for comparison.  The
    log-log plot reveals a power law with an interaction dependent exponent
    $\eta=2K$. (b) $S^{xx}(q=0)/L$ as a function of $1/L$, demonstrating the
    absence of long-range order. Lines are fits to the form
    $S^{xx}(q=0)/L=b/L + c/L^{\eta}$, with $\eta$ taken from (a). Data in (a)
    and (b) are extrapolated to $\theta=\infty$.  (c) Luttinger liquid
    parameters $K$ and $v$. $K$ is extracted from fits to the form
    $S^{xx}(r)=a/r^{\eta}$ shown as lines in (a). $v$ is estimated from the
    density structure factor $N(q=2\pi/L,\tau)$ and finite-size
    extrapolation; $\vF=2\lambda$. }
\end{figure}

\begin{figure}[t]
  \includegraphics[width=0.45\textwidth]{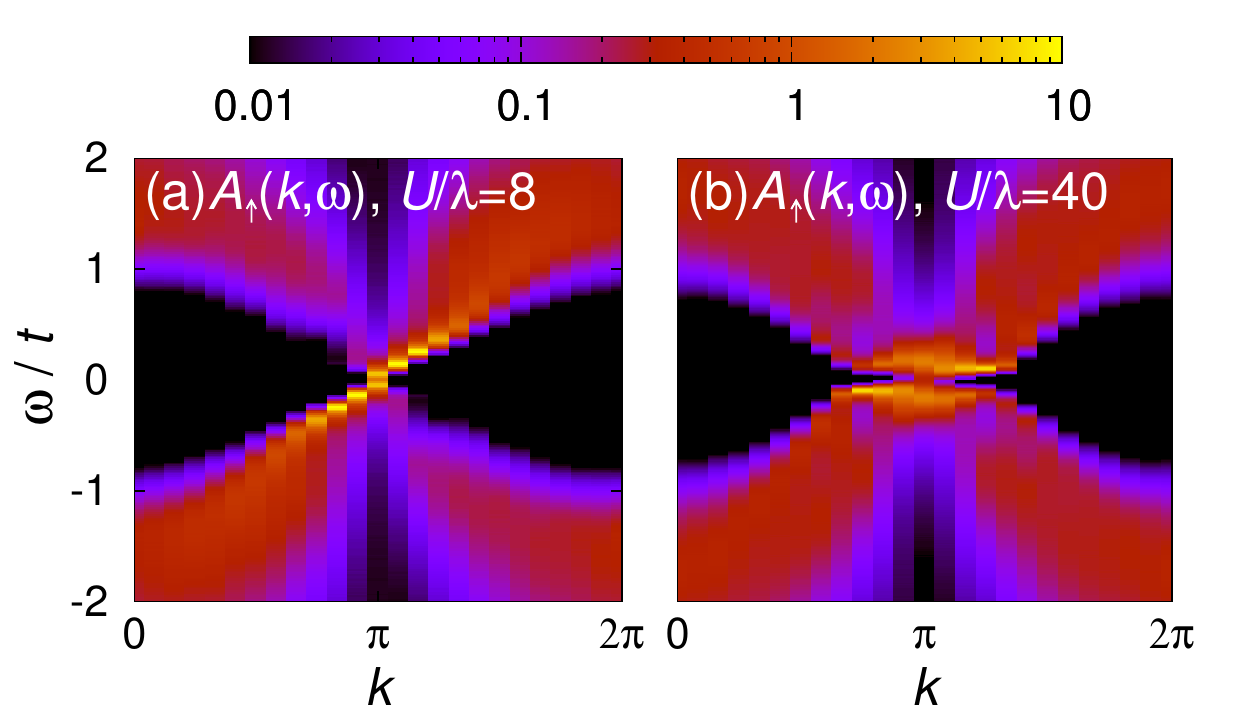}
  \caption{\label{Spectral_function.fig} (Color online)
    Single-particle spectral function $A_\uparrow(k,\omega)$ from
    projector-CTQMC simulations for $L=24$, $\theta=80$ and (a)
    $U/\lambda=8$, (b) $U/\lambda=40$. }
\end{figure}

{\it Results.}---For a zigzag ribbon, the spectral function
$A_\sigma(k,\om)=-\pi^{-1} \text{Im}\,G^\si_{0}(k,\om)$ of the $U=0$ KM model
features a bulk energy gap $\Delta_\text{SO}\sim3\sqrt{3}\lambda$ at the Dirac points, and a
pair of helical edge states with Fermi velocity $\vF\sim2\lambda$ crossing at
$\kF=\pi$ (for half filling) \cite{KaMe05a} [see also
Fig.~\ref{Spectral_function.fig}(a)]. The right (left) movers have spin up
(down), and TRI implies $A^{\uparrow}(k,\omega) = A^{\downarrow}(-k,
\omega)$. The degeneracy at $k=\pi$ is protected by Kramers' theorem.

Correlation effects on energy scales much smaller than the bulk gap can be
studied using bosonization.  In the continuum limit, the fermion operator
becomes $ \Psi_{\sigma}(r) = e^{i \kF x }\mathcal{R}_{\sigma}(x)
\delta_{\sigma,\uparrow} + e^{-i \kF x } \mathcal{L}_{\sigma}(r)
\delta_{\sigma,\downarrow} $.  For a Hubbard interaction, the bosonized
Hamiltonian (the HTL model) reads \cite{Nagaosa99,Wu06,Cenke06} 
\begin{equation}\label{Luttinger_Liquid_H.eq}
  H    
  =  
   \frac{v}{8 \pi}  
  \int_{0}^{L} d x 
  \left\{  \frac{1}{K} [\partial_x \theta_{+}(x)]^2 +  K [\partial_x \theta_{-}(x)]^2 \right\}
  \,.
\end{equation}
Here $-\partial_x \theta_{\pm}(x) /2\pi = \rho_{R}(x) \pm \rho_{L}(x)$, with
$ \rho_{R}(x)$ and $ \rho_{L}(x)$ being the density of right and left movers,
respectively, ${v = \sqrt{\vF^2 -
    (U/2 \pi)^2}}$ and $K = \sqrt{ (\vF - U/2\pi)/(\vF + U/2 \pi)}$.
Single-particle spin-flip scattering between Kramers degenerate states is
blocked by TRI \cite{KaMe05a,Wu06,Cenke06}.  Umklapp processes such as $ e^{i
  4 \kF x} \mathcal{R}^{\dagger}(x) \mathcal{R}^{\dagger}(x+a)
\mathcal{L}(x)\mathcal{L}(x+ a) $ are generally allowed at half filling since
$4\kF = 4 \pi$, but are forbidden here due to spin-$z$ conservation reflected
in the $U(1)$ spin symmetry of the KMH model. Hence, only forward scattering
is left.  The quadratic Hamiltonian~(\ref{Luttinger_Liquid_H.eq}) gives
\begin{align}\label{Luttinger_liquid_corr.eq} \nonumber
  S^{xx}(x)& =\las S^{x}(x) S^{x}(0) \ras \sim \frac{1}{x^{2K}} \cos(2\kF x
  )\,,\\\nonumber
  S^{zz}(x)& =\las S^{z}(x)  S^{z}(0) \ras \sim  \frac{1}{x^2}\,, \\
  N(x) &= \las n(x) n(0) \ras \quad \sim \frac{1}{x^2}\,.
\end{align}
\begin{figure}[t]
  \includegraphics[width=0.45\textwidth]{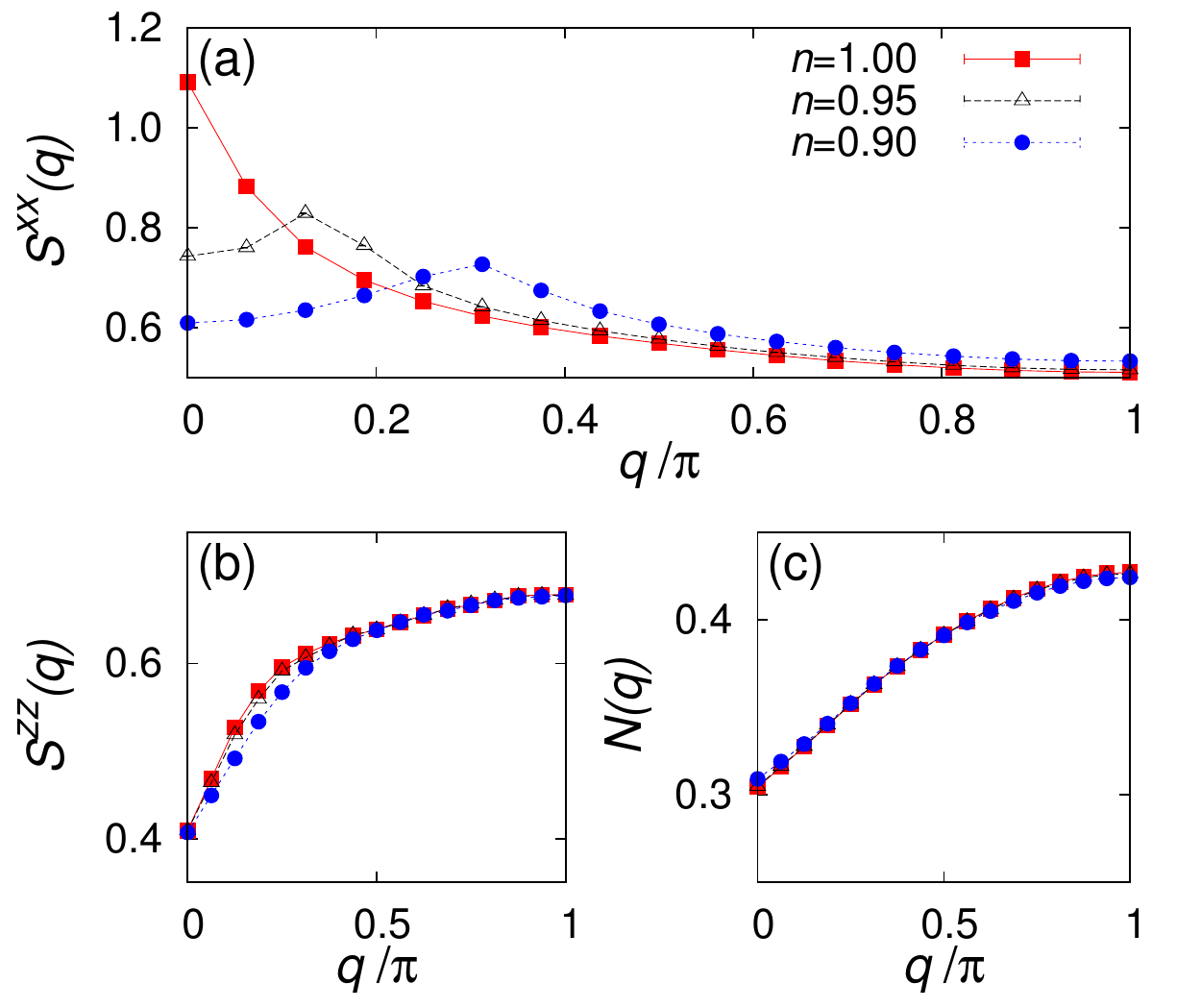}
  \caption{\label{doping.fig} (Color online) Doping dependence of spin and
    charge structure factors for $U/\lambda=8$ from grand-canonical CTQMC
    simulations ($L=32$, $\beta=60$). The $q=0$ points in (b) and (c) are
    linear extrapolations.}
\end{figure}
The transverse correlator $S^{xx}(x)$ involves spin-flip scattering and hence
picks up $2\kF$ oscillations in addition to an interaction dependent
power-law exponent.  $N(x)$ and $S^{zz}(x)$ involve
scattering processes only within the left or right movers, and retain their
Fermi liquid form. For $K<1$, transverse spin correlations dominate.

For $U=0$ ($K=1$), Eq.~(\ref{Luttinger_liquid_corr.eq}) can be verified by
explicitly solving the effective model of Eq.~(\ref{eq:action}).  For $U>0$,
we calculate the correlation functions~(\ref{Luttinger_liquid_corr.eq})
exactly using QMC, and we show $S^{xx}(r)$ at half filling in
Fig.~\ref{Equal-time-correlations.fig}(a). With increasing $U/\lambda$, we
observe a progressively slower decay, corresponding to a reduction of the
exponent $2K$ [see Eq.~(\ref{Luttinger_liquid_corr.eq})]. Similar behavior was
observed in numerical work \cite{PhysRevLett.106.100403,Zh.Wu.Zh.11}. The
finite-size scaling of the structure factor $S^{xx}(q=0)$ shown in
Fig.~\ref{Equal-time-correlations.fig}(b) confirms the absence of long-range,
ferromagnetic order even for $U/\lambda=40$. This conclusion is based on
simulations of very large systems (up to $24\times64$), whereas smaller sizes
would incorrectly suggest long-range order
\cite{Zh.Wu.Zh.11,PhysRevLett.107.166806}. True (Ising) long-range order
becomes possible (at $T=0$) if the $U(1)$ symmetry of the KMH model is further
reduced to $Z_2$ by Rashba coupling. On the mean-field level, symmetry
breaking occurs for any $U>0$ due to a logarithmic instability.

Figure~\ref{Equal-time-correlations.fig}(c) shows the LL parameter $K$, as
obtained from $\eta=2K$ [Eq.~(\ref{Luttinger_liquid_corr.eq})], and the
renormalized velocity $v$ calculated from $N(q,\tau)$ (see below).  $v$
closely follows the Fermi velocity $v_\text{F}$, with a small offset
independent of $\lambda$. This result, consistent with $v$ being inherited
from the bulk and slightly renormalized by $U$, conflicts with
Eq.~(\ref{Luttinger_Liquid_H.eq}). We find $K<1/2$ for the values of
$U/\lambda$ considered, far from the noninteracting limit $K=1$ and in the
regime where umklapp scattering is relevant
\cite{Wu06,Zh.Wu.Zh.11,PhysRevLett.107.166806}.  However, for translationally
invariant systems, this term is allowed only for the special case of exactly
half filling and in the presence of Rashba coupling. 

The $\cos (2\kF r)$ contribution to $S^{xx}(r)$ becomes apparent upon doping
the system by varying the chemical potential inside the bulk gap. As shown in
Fig.~\ref{doping.fig}(a) for $U/\lambda=8$, the peak in the structure factor
$S^{xx}(q)$ moves from $q=0$ to finite $q$ with increasing doping, reflecting
the change of $\kF$.  In contrast, the spin-$z$ and charge structure factors
retain their cusp structure in the long-wavelength limit related to the
$1/r^2$ decay in real space. Deviations from
Eq.~(\ref{Luttinger_liquid_corr.eq}) are observed for $S^{zz}(q)$ outside the
long-wavelength limit for large $U/\lambda\gtrsim10$, and can be related to
the strong-coupling effects discussed below.

Figure~\ref{Spectral_function.fig} shows the spectral function
$A_\UP(k,\om)$.  For weak coupling [$U/\lambda=8$,
Fig.~\ref{Spectral_function.fig}(a)], we see a dominant linear mode crossing
the Fermi level, and the spectrum closely resembles the case $U=0$
\cite{KaMe05a}. With increasing correlations [$U/\lambda=40$,
Fig.~\ref{Spectral_function.fig}(b)], spectral weight is suppressed at low
energies in favor of new high-energy features. The spectral function in
Fig.~\ref{Spectral_function.fig}(b) is fundamentally different from
Fig.~\ref{Spectral_function.fig}(a) and previous results
\cite{PhysRevLett.106.100403}. The latter were qualitatively fully captured
by the HTL model which gives a branch cut $ \langle
\mathcal{R}^{\dagger}(x,t) \mathcal{R}(0,0) \rangle \sim \left( {x - v t}
\right) ^{-\frac{1}{2}(K + 1/K) } $.  These additional features arise only in
the limit of small spin-orbit coupling (large $U/\lambda\gg1$).

The HTL model~(\ref{Luttinger_Liquid_H.eq}) does not account for the
single-particle spectrum in Fig.~\ref{Spectral_function.fig}(b).  Except for
the bulk gap at the Dirac points, the locus of spectral weight at high
energies near $\kF$ bears remarkable resemblance to the zigzag edge states of
graphene ribbons with $U>0$ \cite{PhysRevLett.106.226401}. The physics of the
latter is dominated by quasi-long-range transverse spin fluctuations at the
edge. Similarly, a very slow decay of transverse spin fluctuations at large
$U/\lambda$ [Fig.~\ref{Equal-time-correlations.fig}(a)] is a generic
signature of the strong-coupling regime of the KMH model, see also
Refs.~\cite{PhysRevLett.106.100403,Zh.Wu.Zh.11}.

\begin{figure}[t]
  \includegraphics[width=0.45\textwidth]{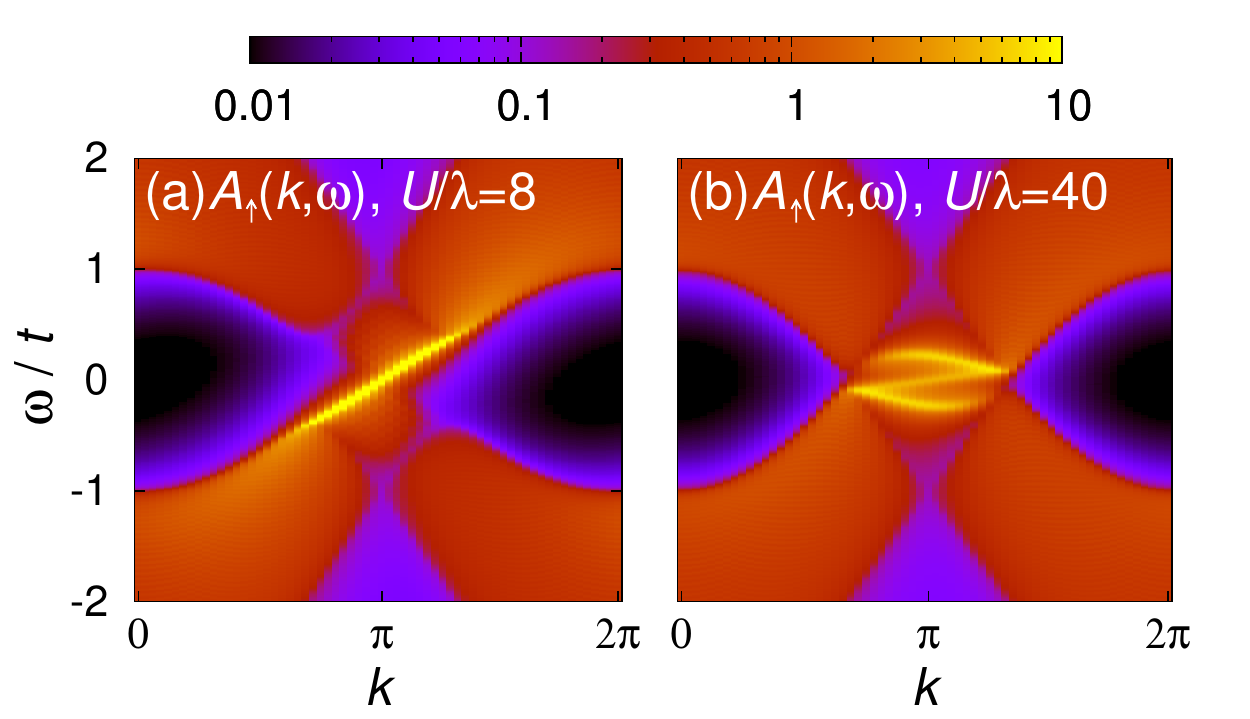}
  \caption{ \label{Perturbation.fig} (Color online) Single-particle spectral
    function $A_\UP(k,\om)$ from second-order perturbation theory. $U/t=2$.}
\end{figure}

To identify spin-flip scattering as the physical origin of the high-energy
features of Fig.~\ref{Spectral_function.fig}(b), we complement our numerical
results by a simple yet sufficient analytical approximation. Rewriting the
Hubbard term as $H_U = - \oh U\sum_{q} \left( S^{+}_q S^{-}_q + S^{-}_q
  S^{+}_q \right)$ with $ S^{+}_q = L^{-1/2} \sum_{k} c^{\dagger}_{k\uparrow}
c_{k+q\downarrow}$, perturbation theory to order $U^2$ gives the self-energy
\begin{equation*}
  \Sigma_{\uparrow}(k,i\omega_m) = \frac{U^2}{\beta L}
  \sum_{q, i \Omega_m} \chi^{\pm}_{0}(q,i \Omega_m)  G^{\downarrow}_0(k-q,i
  \omega_m - i\Omega_m)\,,
\end{equation*}
with inelastic spin-flip scattering off $q=2\kF$ transverse spin fluctuations
described by the susceptibility
%
%\begin{equation*}
 {$ \chi^{\pm}_{0}(q,i \Omega_m)  =
  -
  %\frac{1}{\beta L} 
  (\beta L)^{-1}
  \sum_{k, i \omega_m} G^{\uparrow}_0({k+q},i \omega_m +
  i\Omega_m)   \, G^{\downarrow}_0(k,i \omega_m) $}.
%\,.
%\end{equation*}
%
The results for $A_\UP(k,\om)$ shown in Fig.~\ref{Perturbation.fig}
qualitatively reproduce the numerical data in
Fig.~\ref{Spectral_function.fig}.  For $U/\lambda=40$, the linear low-energy
mode is better visible than in the numerical spectrum whose calculation
involves analytical continuation. The emergence of high-energy features
predominantly above the bulk gap with increasing $U/\lambda$ causes, via the
sum rule $\int d \omega A_{\sigma}(k,\omega) = 1 $, a depletion of spectral
weight at low energies. Spin-flip scattering is present for any $U>0$, but
its effects become apparent in $A_\UP(k,\om)$ when $U/\lambda\gg1$.

\begin{figure}[t]
  \includegraphics[width=0.459\textwidth]{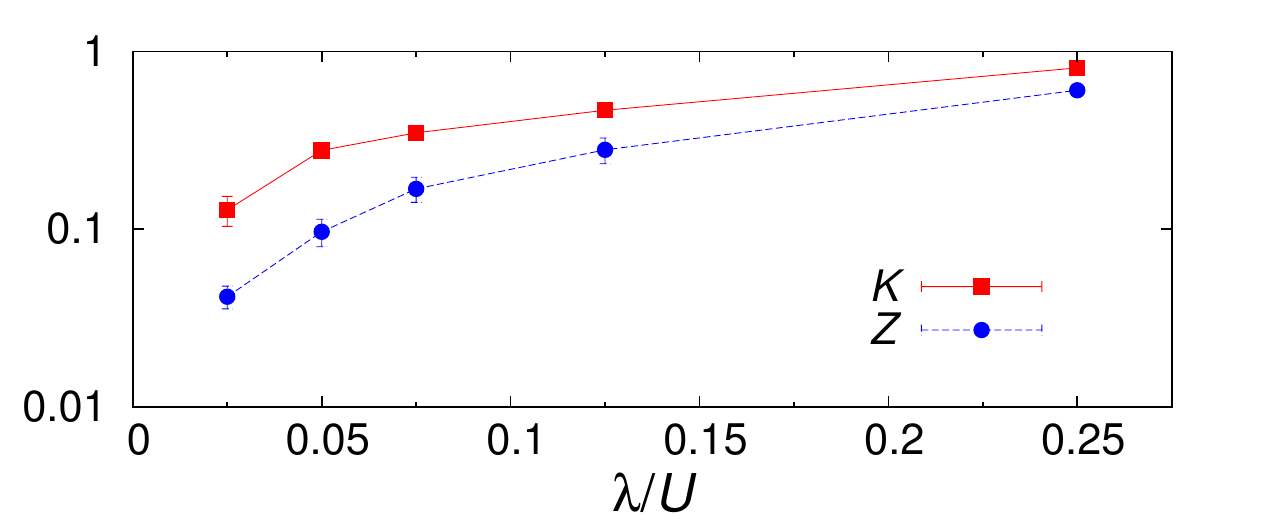}
  \caption{\label{Charge_dyn.fig} (Color online) Luttinger parameter $K$ (from
    Fig.~\ref{Equal-time-correlations.fig}) and spectral weight $Z$ from fits
    to the density structure factor $N(q,\tau)$ at $\theta=80$ and
    finite-size scaling. The  Luttinger liquid result $K=Z$ is violated for a wide range of
    parameters. }
\end{figure}

The deviations from the HTL model and their effect on edge transport may be
quantified by measuring the spectral weight $Z$ of low-energy particle-hole
excitations using the dynamic charge structure factor $N(q,\om)$.  Fitting
the linear mode to the form $N(q,\om) = \oh Z q \delta(\om-vq)$, with weight
$Z$ and velocity $v$ estimated from $N(q=2\pi/L,\tau)$, we can use the
continuity equation to obtain the Drude weight $D= Zv$. This result can be
compared to the relation $D=Kv$ following from
Eq.~(\ref{Luttinger_Liquid_H.eq}), implying $Z=K$ for the HTL in accordance
with results at $U=0$ (not shown). However, as apparent from the QMC results
in Fig.~\ref{Charge_dyn.fig}, $Z\neq K$ over a wide range of $\lambda/U$, and
$Z=K$ holds only asymptotically in the weak-coupling limit. Hence,
irrespective of the interaction strength, inelastic spin-flip scattering
mediated by bulk states leads to deviations from the predictions of the HTL
model~(\ref{Luttinger_Liquid_H.eq}). This effect gains in magnitude with
increasing magnetic correlations (increasing $U/\lambda$). The suppression of
$Z$ (but with $Z=K$) with increasing $U/\lambda$ (see also
Ref.~\cite{PhysRevLett.106.100403}) can be understood in the framework of LL
theory. Interaction effects beyond the model~(\ref{Luttinger_Liquid_H.eq})
are reflected in the pronounced quantitative difference between $Z$ and $K$.

{\it Conclusions.}---Using quantum Monte Carlo simulations, we have
investigated the validity of the low-energy Tomanaga-Luttinger model for edge
states of a $Z_2$ topological insulator. We found the expected interaction
and doping dependence of spin and charge correlations, with dominant
transverse spin fluctuations but no long-range order.  However, the fact that
helical edge modes cannot strictly be energetically separated from the
insulating bulk has important consequences. The bulk states provide phase
space for inelastic spin-flip scattering. In the weak-coupling limit, this
leads to minor but quantifiable violations from bosonization predictions. For
strong coupling, these scattering processes transfer spectral weight from low
to high energies in the single-particle spectral function, and thereby give
rise to features reminiscent of graphene zigzag ribbons.

We thank J.~Budich, T.~Lang, Z.~Y. Meng, P.~Recher, M.~Schmidt, T.~Schmidt,
B.~Trauzettel, S.-C.~Zhang and C.~Xu for discussions. This research was
supported in part by the National Science Foundation under Grant
No. PHY05-51164. We acknowledge support from DFG Grant No.~FOR1162 and
generous computer time at the LRZ Munich and the J\"ulich Supercomputing
Centre.

%\bibliographystyle{../prsty}
%\bibliography{../refs}

\end{document}